\documentstyle[prl,aps,multicol]{revtex}

\draft

\author{M.~A.~Nielsen}
\title{On the units of bipartite entanglement: \\
  Is sixteen ounces of entanglement always equal to one
  pound?}
\address{Center for Quantum Computer Technology, 
  University of Queensland 4072, Australia}
\date{\today}

\begin{document}

\pagestyle{plain}
\pagenumbering{arabic}

\maketitle

\begin{abstract}
  In a good physical theory dimensionless quantities such as the ratio
  $m_p / m_e$ of the mass of the proton to the mass of the electron do
  not depend on the system of units being used.  This paper
  demonstrates that one widely used method for defining measures of
  entanglement violates this principle.  Specifically, in this
  approach dimensionless ratios $E(\rho) / E(\sigma)$ of entanglement
  measures may depend on what state is chosen as the basic unit of
  entanglement.  This observation leads us to suggest three novel
  approaches to the quantification of entanglement.  These approaches
  lead to unit-free definitions for the entanglement of formation and
  the distillable entanglement, and suggest natural measures of
  entanglement for multipartite systems.  We also show that the
  behaviour of one of these novel measures, the entanglement of
  computation, is related to some open problems in computational
  complexity.
\end{abstract}


\begin{multicols}{2}[]
\narrowtext

\section{Introduction}

%
% general subject
%
Quantum mechanics harbours a rich structure whose investigation and
explication is the goal of quantum information
science\cite{Nielsen00a,Preskill98c}.  Of particular interest is the
development of a quantitative theory of entangled quantum states.
This theory is still a work in progress; see for example
\cite{Bennett96a,Nielsen99a,Horodecki96f,Vedral98a,%
  Vidal99a,Vidal00b,Jonathan99a,Bennett99b,Terhal00a,Dur00a} for a
sample of the ongoing work, and further references.  

%
% what the paper does
%
The present paper is a two-pronged contribution to the theory of
entanglement.  The first goal of the paper, explored in detail in
Section~\ref{sec:units}, is to deliver a critique of a commonly used
approach to the definition of measures of entanglement, identifying
what we refer to as a {\em ratio problem} arising in this approach.
This ratio problem manifests itself in, for example, definitions for
measures of entanglement such as the entanglement of formation and
distillable entanglement.  The second goal of the paper, explored in
detail in Section~\ref{sec:new_proposals}, is to develop approaches to
the quantification of entanglement that do not suffer the ratio
problem.  In particular, we describe three novel approaches to the
quantification of entanglement without the ratio problem arising.
These new approaches lead to methods for defining the entanglement of
formation and distillable entanglement that don't suffer the ratio
problem, and also lead to interesting connections between entanglement
measures and computational complexity.  Section~\ref{sec:conc}
concludes the paper.

%
% outline of the paper
%
Before proceeding to the detailed discussion in
Sections~\ref{sec:units} through~\ref{sec:conc}, it is useful to take
a broad look at how the ratio problem arises, and to consider some of
the issues inherent in attempts to critique definitions of
entanglement.  A central notion in much existing work on the
quantification of entanglement is the concept of a ``standard unit''
of entanglement, usually chosen to be the Bell state of two qubits,
$(|00\rangle+|11\rangle)/\sqrt 2$.  Naturally, one may ask whether the
quantitative theory of entanglement changes in any essential way when
the standard unit of entanglement is changed to be a state other than
the Bell state?  In general, in a good physical theory dimensionless
quantities such as the ratio $m_p / m_e$ of the mass of the proton to
the mass of the electron should not depend on the system of units
being used, whether they be kilograms, Planck masses, or any other
unit.  For convenience, I will refer to this unit-invariance as the
{\em ratio property}.  In this paper we show that the usual method of
defining the entanglement of formation and the distillable
entanglements don't satisfy the ratio property: for both these
measures, ratios $E(\rho)/E(\sigma)$ of the entanglement present in
states $\rho$ and $\sigma$ depend on what standard unit of
entanglement is chosen, in contradiction with our expectations for
well-behaved physical quantities.  This is the ratio problem. 

%
% what the paper does
%
How important is the ratio problem to ongoing attempts to construct a
theory of entanglement?  As explained in more detail below, I believe
considerable caution is warranted in assessing criticisms of measures
of entanglement which are of an axiomatic nature, as is the ratio
problem.  In particular, I don't believe that the ratio problem {\em
  alone} is sufficient to invalidate the notion of defining measures
of entanglement based on a standard unit of entanglement; I think that
this notion can and should be further developed, as is being done in,
for example, \cite{Bennett99c,Linden99c,Vidal00c}.  On the other hand,
I believe that the ratio problem is of sufficient importance that it
should strongly motivate the search for other approaches to defining
measures of entanglement which don't suffer this problem, and some
progress along these lines is reported later in this paper.

%
% dangers 
%
I commented above that caution is necessary in assessing criticisms of
an axiomatic nature.  The reason for this comment is the general
poverty of such axiomatic approaches.  By this, I mean that although
there is a tendency when defining measures of entanglement to focus on
``nice'' elementary properties one would like such a measure to have
(such as the ratio property), what we really want is deep theorems
connecting measures of entanglement in surprising ways to other
problems in quantum information science.  An analogous situation
occurred in the early days of information theory: the Shannon entropy
derives its goodness as a measure of information not from its nice
elementary properties, as described in any information theory text,
but from deep theorems connecting the entropy to {\em a priori}
unrelated problems of data compression, communication in the presence
of noise, and so on.  At the present time I believe it is fair to say
that few deep results connecting measures of entanglement to other
problems are known.  Of course, in the absence of such deep results,
critiques (in particular, this paper) of existing definitions of
entanglement are necessarily of an axiomatic nature, but the above
caveats need to be borne in mind when assessing such critiques.

\section{The ratio problem}
\label{sec:units}

%
% define the entanglement of formation and distillation
%
The entanglement of formation and the distillable entanglement are two
of the best known measures of entanglement for bipartite quantum
systems\cite{Bennett96a}.  We now review the standard definitions of
these measures, based on the Bell state as the standard unit of
entanglement, writing the entanglement of formation for a general
quantum state $\rho$ as $F(\rho)$, and the distillable entanglement as
$D(\rho)$.  In this approach the entanglement of formation is defined
as follows.  Imagine Alice and Bob have a protocol that, given $m$
shared Bell states, allows them to create $n$ ``pretty good'' copies
of $\rho$, using only local operations on their respective systems,
and classical communication --- a so-called ``LOCC'' protocol.  These
copies of $\rho$ must be pretty good in the sense that the fidelity
between $\rho^{\otimes n}$ and the state created by the protocol tends
to $1$ as $m$ and $n$ tend to infinity.  The entanglement of formation
is defined to be the minimum value that may be achieved for the ratio
$m / n$, in the limit as $n$ becomes large, where the minimum is
performed over all possible LOCC protocols.  Schematically, we have:
\begin{eqnarray}
  n F(\rho) \times \mbox{Bell} \rightarrow n \times \rho,
\end{eqnarray}
where the $\rightarrow$ indicates a LOCC protocol.  Thus, the
entanglement of formation is a measure of how cheaply $\rho$ may be
created, given Bell state entanglement and the ability to perform
local operations and classical communication.  

%
% definition of the distillable entanglement
%
In this Bell-state-as-standard-unit approach the distillable
entanglement is defined in a similar asymptotic way.  Specifically,
consider a LOCC protocol in which Alice and Bob start out sharing $n$
copies of $\rho$, and produce $m$ good copies of the Bell state.  The
distillable entanglement is defined as the maximum value that may be
achieved for the ratio $m / n$, in the limit as $n$ becomes large.
Schematically, we have:
\begin{eqnarray}
  n \times \rho \rightarrow n D(\rho) \times \mbox{Bell}.
\end{eqnarray}
Thus, the distillable entanglement is a measure of the Bell state
entanglement yield which may be produced if Alice and Bob start with
copies of $\rho$ and are given the ability to perform local
operations and classical communication.

%
% what is known about the entanglement of formation and distillable
% entanglement
%
An impressive but incomplete body of knowledge exists about the
entanglement of formation and distillable entanglement.  For pure
states $|\psi\rangle$ both quantities are equal to the entropy of
Alice's reduced density matrix\cite{Bennett96c,Bennett96a}, $\rho_\psi
\equiv \mbox{tr}_B(|\psi\rangle \langle \psi|)$, that is,
$F(|\psi\rangle \langle \psi|) = D(|\psi\rangle \langle \psi|) =
S(\rho_\psi)$.  For the case of mixed states of two qubits,
Wootters\cite{Wootters98a} has provided an elegant formula for the
entanglement of formation, subject to the proof of the {\em additivity
  conjecture} for entanglement of formation.  (This conjecture has not
yet been proved, but is widely believed to be true after numerical
testing.)  The distillable entanglement has been evaluated or bounded
for a variety of classes of mixed states (for a selection of work and
further references see
\cite{Rains00a,Shor00a,Eisert00a,Rains98a,Vedral98a}), however no
general formula is known, even for the case of a mixed state of two
qubits.

%
% define entanglement of formation and distillable entanglement with
% respect to a new unit
%
Suppose we change the standard unit of entanglement so it is no longer
the Bell state, defining a new ``entanglement of formation'' and
``distillable entanglement'' with respect to a new standard unit of
entanglement, the state $\sigma$.  We denote these new concepts as
$F_\sigma(\rho)$ and $D_\sigma(\rho)$, and refer to them as the
$\sigma$-entanglement of formation, and the $\sigma$-distillable
entanglement.  More precisely, the $\sigma$-entanglement of formation
is defined as the minimal possible value that may be achieved for the
ratio $m / n$, in the limit as $n$ becomes large, for a LOCC protocol
in which Alice and Bob initially share $m$ copies of $\sigma$, and
produce $n$ pretty good copies of $\rho$, with the same asymptotic
fidelity requirements as before.  The $\sigma$-distillable
entanglement is defined as the maximal possible value that may be
achieved for the ratio $m / n$, in the limit as $n$ becomes large, for
a LOCC protocol in which Alice and Bob initially share $n$ copies of
$\rho$, and produce $m$ pretty good copies of $\sigma$.

%
% proof of the various inequalities
%
With these definitions it is easy to prove the inequalities
\begin{eqnarray} 
  \label{eq:bounds_1}
  D(\rho) \leq F_\sigma(\rho)D(\sigma) \leq F(\rho) \leq
  F_\sigma(\rho)F(\sigma) \\
  \label{eq:bounds_2}
  D_\sigma(\rho) D(\sigma) \leq D(\rho) \leq D_\sigma(\rho) F(\sigma) 
  \leq F(\rho).
\end{eqnarray}
We prove the inequalities in~(\ref{eq:bounds_2}); the inequalities
in~(\ref{eq:bounds_1}) follow by similar reasoning.  In general it
will be convenient to concentrate the discussion on $D_\sigma(\rho)$,
since the properties of $F_\sigma(\rho)$ are similar.  By definition
of the distillable entanglement and entanglement of formation, there
exist asymptotically good LOCC protocols achieving the transformations
$n \times \rho \rightarrow nD(\rho) \times \mbox{Bell}$ and $nD(\rho)
\times \mbox{Bell} \rightarrow nD(\rho)/F(\sigma) \times \sigma$, and
thus there is an asymptotically good protocol achieving the
transformation $n \times \rho \rightarrow n D(\rho)/F(\sigma) \times
\sigma$.  By definition of the $\sigma$-distillable entanglement, it
follows that $D_\sigma(\rho) \geq D(\rho)/F(\sigma)$, which is the
second inequality in~(\ref{eq:bounds_2}).  To prove the third
inequality, note that by definition of the $\sigma$-distillable
entanglement there exists an asymptotically good LOCC protocol $n
\times \rho \rightarrow nD_\sigma(\rho) \times \sigma$.  By definition
of the entanglement of formation, there exists an asymptotically good
LOCC protocol $n F(\rho) \times \mbox{Bell} \rightarrow n \times
\rho$, and thus there exists an asymptotically good LOCC protocol $n
F(\rho) \times \mbox{Bell} \rightarrow nD_\sigma(\rho)$, from which it
follows that $F(\sigma) \leq F(\rho)/D_\sigma(\rho)$, which is the
third inequality in~(\ref{eq:bounds_2}).  The first inequality follows
by noting that there are asymptotically good LOCC protocols $n \times
\rho \rightarrow nD_\sigma(\rho) \times \sigma \rightarrow
nD_\sigma(\rho) D(\sigma) \times \mbox{Bell}$, and thus $D(\rho) \geq
D_\sigma(\rho)D(\sigma)$.

%
% evaluation at \sigma
%
Suppose $\sigma$ is a state whose distillable entanglement is strictly
less than the entanglement of formation, $D(\sigma) < F(\sigma)$.
Such states have been proved to exist in \cite{Horodecki00b}.
Intuitively it is clear that $D_\sigma(\sigma) = 1$, and we prove this
rigorously below.  From the definition it follows immediately that
$D_\sigma(\mbox{Bell}) = 1/F(\sigma)$, and thus
\begin{eqnarray}
\frac{D_\sigma(\sigma)}{D_\sigma(\mbox{Bell})}
 = F(\sigma) > D(\sigma) = \frac{D(\sigma)}{D(\mbox{Bell})},
\end{eqnarray}
providing an example where dimensionless ratios of entanglement
quantities depend upon the unit of entanglement being used.  To
conclude, we need only show that $D_\sigma(\sigma) = 1$.  Clearly,
$D_\sigma(\sigma) \geq 1$; we need only worry that it might be
possible for Alice and Bob to distill more than one copy of $\sigma$
per copy of $\sigma$ that they are given.  (Note that this {\em is}
possible when $\sigma$ is separable, since additional copies of
$\sigma$ can be prepared by LOCC.)  For a contradiction, suppose that
$D_\sigma(\sigma) > m/n > 1$, where $m$ and $n$ are large positive
integers.  Then in the asymptotic limit of large $m$ and $n$, we are
able to accomplish the transformation $\sigma \times n \rightarrow
\sigma \times m$ with high fidelity.  Assuming the additivity
conjecture for entanglement of formation, and using the continuity
results of Nielsen\cite{Nielsen00b} and the fact that the entanglement
of formation is non-increasing under LOCC, we obtain $n F(\sigma) \geq
m F(\sigma)$, that is, $n \geq m$, the desired contradiction.

%
% case for F_\sigma
%
Similar results are easily obtained for the $\sigma$-entanglement of
formation.  Specifically, suppose $D(\sigma) < F(\sigma)$.  It is
easily verified that $F_\sigma(\sigma) = 1$ and $F_\sigma(\mbox{Bell})
= 1/D(\sigma)$, so
\begin{eqnarray}
  \frac{F_\sigma(\sigma)}{F_\sigma(\mbox{Bell})} = D(\sigma) <
  F(\sigma) = \frac{F(\sigma)}{F(\mbox{Bell})},
\end{eqnarray}
showing that the ratio problem arises for both the distillable
entanglement and the entanglement of formation.

%
% possible objection
%
We have shown that the usual definitions for the entanglement of
formation and distillable entanglement do not satisfy the ratio
property.  In the introduction some remarks were made about how
seriously such a problem needs to be taken.  Further to those remarks,
the ratio problem could possibly be disregarded on the grounds that it
is ``unnatural'' to choose a mixed state as the standard unit of
entanglement, or that the Bell state really is, in some sense,
special.  I don't believe that such a position is valid if one takes
seriously the notion that quantifying the amount of entanglement
present in a mixed state is a sensible objective; surely if this is
the case then it makes sense to change the unit of entanglement which
one uses, and expect dimensionless ratios to remain invariant under
such changes in the unit.  To continue the analogy with mass: while
there are serious practical concerns of stability, repeatability and
ease-of-duplication when using poorly chosen standards of mass (say,
one mass unit $\equiv$ one Indian elephant), from an ideal theoretical
viewpoint this is just as satisfactory as using, say, electrons as the
mass standard.  
%Note also that if one adopts this point of view then
%the existence of bound entanglement is really a consequence of what
%system of units is adopted, for $D_\sigma(\sigma) = F_\sigma(\sigma) =
%1$ for bound entangled states.

\section{Novel approaches to the definition of entanglement}
\label{sec:new_proposals}

%
% other possible approaches: need for an approach that is not units
% dependent, and mention the relative entropy of entanglement
%
The ratio problem suggests that it might be useful to develop
approaches to quantifying entanglement which do not suffer this
problem.  Some such approaches already been proposed in the existing
literature, such as the relative entropy of
entanglement\cite{Vedral98a}, which avoids the ratio problem by not
singling out any special standard unit of entanglement --- they are
``unit-free''.  We now describe several new methods for defining
entanglement which do not suffer the ratio problem.  Furthermore, we
will show that these definitions give rise to unit-free definitions of
the entanglement of formation and distillable entanglement,
definitions which do not suffer the ratio problem!  What then have we
gained by switching to new definitions?  The answer is two-fold.
First, from a fundamental point of view, definitions of entanglement
which do not satisfy the ratio problem are {\em a priori} more
satisfactory.  Second, these alternative definitions generalize in a
natural way to multipartite systems, and we will see that they have
natural connections to problems in quantum computational complexity,
quantum communication, and distributed quantum computation.

%%
%% entanglement of creation
%%
Our first alternative approach to the definition of an entanglement
measure is the {\em entanglement of creation}, $E_{\rm{cr}}(\rho)$,
defined as follows.  Alice and Bob wish to create some large number
$n$ of good copies of the state $\rho$.  They start with no preshared
entanglement, but can use qubits to communicate (at a cost), and can
do classical communication for free.  The entanglement of creation is
defined to be the asymptotically minimal value for the ratio of the
number of qubits $m$ they transmit to the number of copies $n$ of
$\rho$ they create, with the requirement that as $n$ becomes large the
copies of $\rho$ created must be good copies.  This definition is
obviously independent of any concept of a standard unit of
entanglement, and thus can not possibly suffer from the ratio problem.
Note that we {\em do} use a specific ``standard'' quantum channel for
Alice and Bob to communicate --- a noiseless quantum channel ---
however it seems clear that this channel really is special among all
possible communication channels.

%
% equivalence with the entanglement of formation
%
With this definition, the entanglement of creation is equal to the
entanglement of formation, by the following argument.  Suppose we are
given a protocol for creating $n$ copies of $\rho$ using
$nE_{\rm{cr}}(\rho)$ qubits of communication between Alice and Bob,
plus classical communication.  Then this is easily simulated using
$nE_{\rm{cr}}(\rho)$ shared Bell states, classical communication, and
no quantum communication, using quantum
teleportation\cite{Bennett93a}.  Thus, $F(\rho) \leq
E_{\rm{cr}}(\rho)$.  Conversely, suppose we are given a protocol to
create $n$ copies of $\rho$ using $nF(\rho)$ shared Bell states, local
operations and classical communication.  This may be simulated using
$nF(\rho)$ qubits of communication to set up the $nF(\rho)$ shared
Bell states, and classical communication.  Thus, $E_{\rm{cr}}(\rho)
\leq F(\rho)$, from which we deduce that $E_{\rm{cr}}(\rho) =
F(\rho)$, as claimed.

%%
%% entanglement of communication
%%
Our second alternative approach to the definition of entanglement is
the {\em entanglement of communication}.  Suppose Alice and Bob share
a large number $n$ of copies of a state $\rho$.  The entanglement of
communication $E_{\rm{comm}}(\rho)$ of $\rho$ is defined to be the
asymptotically maximal value for the ratio $m / n$ of the number of
qubits $m$ Alice can transmit to Bob with asymptotically high
fidelity, using LOCC, and $n$ preshared copies of $\rho$.  With this
definition the entanglement of communication is exactly equal to the
distillable entanglement, as may be seen by the following argument.
By definition, there exists a LOCC protocol for Alice and Bob to
convert $n$ copies of $\rho$ into $nD(\rho)$ copies of the Bell state.
Then using teleportation\cite{Bennett93a} and additional classical
communication Alice can transmit $nD(\rho)$ qubits to Bob.  Thus
$D(\rho) \leq E_{\rm{comm}}(\rho)$.  To show the reverse inequality,
suppose Alice and Bob share $n$ copies of $\rho$.  Alice creates $n
E_{\rm{comm}}(\rho)$ Bell states locally, and then transmits half of
each Bell state to Bob, creating $n E_{\rm{comm}}(\rho)$ shared Bell
states, and establishing that $E_{\rm{comm}}(\rho) \leq D(\rho)$, from
which we deduce that $E_{\rm{comm}}(\rho) = D(\rho)$, as claimed.  As
for the entanglement of creation, the entanglement of communication is
obviously independent of any concept of a standard unit of
entanglement, and thus can not possibly suffer from the ratio problem.

%
% review these two approaches
%
As we have shown, the entanglement of creation and entanglement of
communication are not truly new measures of entanglement.  Rather,
they offer unit-free methods for defining the quantity of entanglement
present in a system.  A significant advantage of both methods of
definition over the standard approach is that they suggest natural
generalizations to multipartite systems, where it is not at all clear
what the appropriate standard unit of entanglement is.  For example,
for an $n$ party system, it is possible to define the entanglement of
creation for a state $\rho$ to be the minimal asymptotic amount of
qubit communication between the parties, per good copy of the state
created.  In the case of the entanglement of communication, making
appropriate definitions is a little more tricky; however, as the
following example illustrates, there are some natural starting points.
One possible way to go is to define the entanglement of communication
for a state $\rho$ as the {\em total} number of qubits of
communication which can be done (between all possible pairs of
parties), per copy of $\rho$.  Other natural variants of this measure
also suggest themselves, perhaps counting the number of qubits of
communication that can be achieved between some specified pair of
parties, or some other natural communication goal.  These measures of
entanglement will be investigated in more detail elsewhere, however
for now the main point is that they provide illustrative examples of
alternative approaches to the definition of the entanglement of
formation and distillable entanglement which do not suffer the ratio
problem, and yield useful generalizations.

%
% entanglement of computation
%
Our third and final alternative approach to the quantification of
entanglement is the {\em entanglement of computation}, which is best
illustrated via an example.  Suppose we set $U$ to be the controlled
phase gate\cite{Nielsen00a} on two qubits, that is, the gate which
takes $|x,y\rangle$ to $(-1)^{xy} |x,y\rangle$.  The idea is that
Alice and Bob are able (at a cost) to apply the gate $U$ jointly to
one of Alice's qubits and to one of Bob's qubits.  We also allow them
to communicate classically for free; what happens when free classical
communication is not allowed will be investigated elsewhere.  The
entanglement of computation for $\rho$ with respect to $U$,
$E_U(\rho)$, is defined to be the minimal asympotic value for the
ratio $m/n$ of the number of times $m$ the gate $U$ must be applied to
create $n$ pretty good copies of the state $\rho$.  Obviously, there
is no standard unit of entanglement invoked in this definition, and in
this sense the entanglement of computation does not suffer the ratio
problem.  Of course, we did single out a standard gate, $U$, in our
definition, and we will ask below whether this might cause some
analogue of the ratio problem to arise for the entanglement of
computation.  In the specific case where $U$ is the controlled phase
gate, it is easily seen that $E_U(\rho) = F(\rho))$.  This follows
from the results of \cite{Collins00a}, where it was shown that, given
the ability to communicate classically, one shared Bell state is
equivalent to the ability to perform a controlled-{\sc not} (or
equivalently, a controlled phase) gate.

%
% difficulties
%
We have seen that for a specific choice of gate $U$ the entanglement
of computation is equal to the entanglement of formation, however in
general it is not obvious that this is the case.  Indeed, it is
conceivable that ratios like $E_U(\rho) / E_U(\sigma)$ may depend on
the quantum gate $U$ chosen as the standard unit, creating a sort of
ratio problem for the entanglement of computation.  I conjecture that
such a situation does in fact occur, but have not yet proved it; this
issue will be explored in more detail elsewhere.  Furthermore, it is
not obvious that the definition is symmetric under interchange of the
role of Alice and Bob; this is true in the case when $U$ is chosen to
be the controlled phase gate, however for gates without the symmetry
of the controlled phase gate it is not so obvious.

%
% point out the advantage that it can easily be generalized to n
% qubits.  Conjecture that it is exponential in n, in general.
%
Despite these possible difficulties a major advantage of the
entanglement of computation is that it generalizes to multipartite
systems in an obvious way.  For example, for a system consisting of
$n$ qubits, we can define the $n$ party entanglement of computation
for a state $\rho$, with respect to the controlled phase gate, to be
the number of controlled phase gates required, asymptotically, to
produce a copy of $\rho$, with local operations on individual qubits
and classical communication allowed.  It follows from results of
Knill\cite{Knill95a} that for most states of $n$ qubits this measure
of entanglement is, in general, exponential in the number of qubits
$n$ in the system.

%
% connection to computational complexity
%
This generalization to $n$ qubits allows us to deduce an interesting
connection between the entanglement of computation and computational
complexity theory, using a method similar to that developed in
Chapter~6 of \cite{Nielsen98d}, where entanglement measures were used
as a tool to deduce bounds on distributed quantum computation.
Suppose we consider the state
\begin{eqnarray}
  |\psi_n\rangle \equiv \sum_G |G\rangle |f(G)\rangle,
\end{eqnarray}
where the sum is over all graphs $G$ on $n$ vertices, $|G\rangle$ is
some appropriate representation for $G$ in the computational basis,
such as a listing of the elements of the adjacency matrix for $G$, and
$f(G) \in \{ 0,1 \}$ is some binary function of $G$.  For example, we
might define $f(G) \equiv 0$ if $G$ has no Hamiltonian cycle, while
$f(G) \equiv 1$ if $G$ has a Hamiltonian cycle.  Suppose we have a
quantum algorithm to compute $f(G)$ in time $g(n)$.  It follows that
the entanglement of computation for $|\psi_n\rangle$ must scale as
$O(g(n))$, with some small overhead necessary to prepare the
superposition $\sum_G |G\rangle$.  Suppose it were possible to prove
that for this choice of function $f$, the entanglement of computation
for the state $|\psi_n\rangle$ behaved in a manner superpolynomial in
$n$\footnote{Strictly speaking, an extra system containing working
  qubits should be appended to the state $|\psi_n\rangle$ to make this
  line of argument rigorous.  Without loss of generality we may assume
  that the working qubits are in the all $|0\rangle$ state.}.  Then it
would follows that $\mathbf{QP}$, the class of languages accepted with
probability one on a quantum Turing machine operating in polynomial
time, is not equal to the class $\mathbf{NP}$ of languages accepted by
a classical non-deterministic Turing machine.  This would further
imply that $\mathbf{P} \neq \mathbf{NP}$, settling one of the great
open problems of theoretical computer science.  Thus, the problem of
determining the quantity of entanglement present in a quantum state
are of relevance to questions about computational complexity theory.
(It is interesting to note that similar lines of thought can be
explored in relation to the complexity class $\mathbf{BQP}$, although
additional continuity properties are required for entanglement
measures in this case.)  For this line of attack on questions about
computational complexity I believe it is quite encouraging that the
entanglement of computation can be calculated exactly in the two qubit
case\cite{Wootters98a}, and for calculation in the $n$ qubit case our
large body of knowledge concerning asymptotic behaviour (especially
various laws of large numbers) is encouraging, as is recent progress
on classifying the orbits of the $n$ qubit local unitary group (see
for example \cite{Carteret00a,Acin00a,Linden98a,Grassl98a}, and
references therein).

%
% caveat
%
A caveat about this approach to problems in computational complexity
is in order.  Formally, the state $|\psi_n\rangle$ is rather similar
to a classical probability distribution on ordered pairs $(G,x)$,
where $G$ is an $n$ vertex graph and $x$ a single bit, defined by
\begin{eqnarray}
p(G,x) \equiv \left\{ \begin{array}{l} 1/M \mbox{ if } x = f(G) \\
   0 \mbox{ if } x \neq f(G), \end{array} \right.
\end{eqnarray}
where $M$ is the total number of graphs on $n$ vertices.  Once could
define a notion of ``computational correlation'' measuring the amount
of correlation in this probability distribution, analogous to the
entanglement of computation, except using classical gates and
randomness to create the probability distribution.  Results about
computational complexity would then follow from results about the rate
of growth of the computational correlation of the distribution $p$ as
$n$ grows, in a fashion exactly analogous to the argument outlined
above for the entanglement of computation.  This is, perhaps, an
approach that could be taken to open problems in computational
complexity; nonetheless, I believe that there is more hope in the
quantum approach, since quantum information theory has arguably a more
elegant structure than classical information theory, as indicated for
example by Wootters' beautiful work\cite{Wootters98a} on the
entanglement of formation (and thus entanglement of computation) for
two qubits.

\section{Conclusion}
\label{sec:conc}

%
% conclusions: stress the need for more, and more non-trivial,
% connections of work on entanglement to other basic
% tasks like computation and communication
%
The present paper has focused on the axiomatics of entanglement.  The
problems addressed are elementary, but fundamentally influence how one
thinks about generalizations of entanglement measures to multipartite
systems.  In particular, I believe that the approach to entanglement
measures based upon the notion of a standard unit of entanglement,
while fruitful, needs to be supplemented by other approaches to making
such definitions, such as the relative entropy of entanglement, and
the three approaches suggested here.  At present it is not clear what
approach to the quantification of entanglement will ultimately yield
the most insight, so it is crucial to think critically about existing
approaches, and to develop conceptually novel approaches to the
quantification of entanglement.

\section*{acknowledgments}
Thanks to Carl~Caves, Julia~Kempe and Damian~Pope for stimulating
discussions about entanglement.

%\bibliography{../../../mybib}

\end{multicols}

\end{document}